\documentclass{emulateapj}
\usepackage{psfig}
\usepackage{natbib}
\shorttitle{Far-UV H$_{2}$ in Orion}
\shortauthors{France \& McCandliss}

\begin{document}


\title{Molecular Hydrogen in Orion as Observed by the $Far$ $Ultraviolet$
$Spectroscopic$ $Explorer$}


\author{K. France and S. R. McCandliss}
\affil{Department of Physics and Astronomy, Johns Hopkins University,
    Baltimore, MD 21218}

%\author{P. D. Feldman \altaffilmark{1}}
%\affil{Department of Physics and Astronomy, Johns Hopkins University,
%    Baltimore, MD 21218}

%\and
%\author{Author list?\altaffilmark{N}}
%\affil{Dept, School,
%    Place}


%%%These are for extra affiliations at the bottom of page
%\altaffiltext{1}{Visiting Astronomer, Cerro Tololo Inter-American Observatory.
%CTIO is operated by AURA, Inc.\ under contract to the National Science
%\altaffiltext{2}{present address: Center for Astrophysics,
%    60 Garden Street, Cambridge, MA 02138}
%\altaffiltext{3}{Patron, Alonso's Bar and Grill}

\begin{abstract}
Diffuse far-ultraviolet stellar
emission scattered by dust grains has been observed
in a region near the Orion Nebula.  In addition to the scattered 
stellar continuum,  emission and
absorption features produced by molecular hydrogen
have been identified.  In this Letter, we present an 
analysis of this absorption and fluorescent emission
from molecular hydrogen in Orion.  We model the
spectra obtained with the $Far$ $Ultraviolet$ $Spectroscopic$ $Explorer$
using optical depth templates and a fluorescent emission code.  
These results are surprising because previous studies
have found little ultraviolet absorption from H$_{2}$ in this region, and
the emission is coming from a seemingly empty part of the nebula.
We find that the emission fills in the observed
 absorption lines where the two overlap.
These data support the claim that fluorescent excitation by ultraviolet photons
is the primary mechanism producing the 
near-infrared emission spectrum observed in the outer regions
of the Orion Nebula.
\end{abstract}


\keywords{ISM:molecules~---~ISM:individual~(M 42)~---~
	reflection nebulae~---~ultraviolet:ISM}

%%%%%%%%%%%%%%%%%%%%%%%%%%%START OF THE PAPER%%%%%%%%%%%%%%%%%%%%%%%
\section{Introduction}

The Orion Nebula (M42) is among the most well-studied objects
in the sky.  Its proximity to the Sun combined with a bright Photo-Dissociation Region (PDR) and 
a giant molecular cloud make it an ideal region to study a range of
astrophysical processes.  
Molecular hydrogen (H$_{2}$) should account
for the majority of the molecular mass in this region. 
Observations of emission and absorption features of this molecule provide useful 
diagnostics of the physical 
conditions in PDRs and molecular 
clouds~\citep{burton89,luhman94,rachford02,kristensen03,france05}.  
Previous observations of H$_{2}$ emission in Orion 
have been limited to the weak quadrupole rovibrational transitions arising in the
near and mid-infrared (IR; see references above as well as Habart et al. 2004).
In this Letter, we present an analysis of the first 
far-ultraviolet fluorescent emission spectra of the electronic
transitions of H$_{2}$ in Orion, obtained by the $Far$ $Ultraviolet$
$Spectroscopic$ $Explorer$ ($FUSE$).  These spectra have been presented
by~\citet{murthy05}, in
work focusing on the dust scattering properties of the region.~\nocite{murthy05}
The spectra show a dust scattered continuum, onto which 
absorption and emission features of H$_{2}$ are identified.

Molecular hydrogen can be excited into upper rovibrational levels
via fluorescent excitation by ultraviolet (UV) photons, 
non-thermal electrons, and collisionally (thermally), by shocks.
In a gas-rich region that contains outflows from young stars
and an intense ultraviolet radiation field produced by O and B stars, 
it seems likely that multiple processes are at work.  
Near-IR observations of H$_{2}$ in Orion have been used to distinguish
which of these excitation mechanisms are consistent with the observed emission~\citep{luhman94,kristensen03}.
These $FUSE$ observations do not rule out the possibility of a contribution
from shocks, however they provide solid evidence for UV fluorescence in Orion.

In \S2, we present the $FUSE$ data obtained during a 
serendipitous pointing equilibration observation (see~\citet{murthy05} for a comprehensive description
of the observations).  \citet{murthy05} find that the dust scattered 
light is dominated by the intense radiation
field of $\theta^{1}$ Ori C (HD 37022; O6V~Maiz-Apellaniz et al. 2004),
despite the small angular separation 
of the observed field from HD~36981 (B5V)~\nocite{maiz04}.  
Given its intense far-UV output, 
$\theta^{1}$ Ori C is likely to be the dominant source of 
excitation for the H$_{2}$ emission observed in Orion.
A model of the H$_{2}$ fluorescnce, using the unattenuated continuum of
$\theta^{1}$ Ori C as the excitation source, is described in \S3; that model is  
compared to the observations in \S4.
These observations provide evidence that UV excitation gives rise to the observed
near-IR emission lines in the Orion Nebula region, 
in agreement with the findings of~\citet{luhman94}.

\begin{figure}
\begin{center}
\epsscale{0.8}
\rotatebox{90}{
\plotone{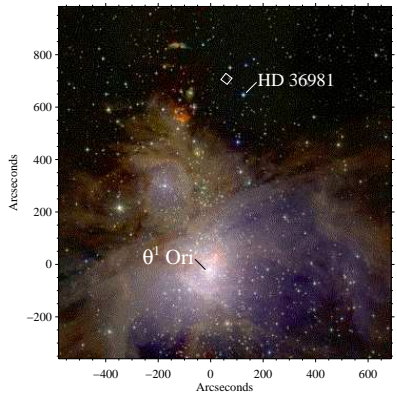} }
\caption{\label{slits} 2MASS Atlas image of the Orion Nebula, M42.  $\theta^{1}$ 
Ori is at the origin and the $FUSE$ LWRS aperture overlay is shown at the upper right
near HD 36981. 
The pointing was determined from an analysis of the FES image 
of the target field.
Atlas Image mosaic obtained as part of the 
Two Micron All Sky Survey, a joint project of the University of 
Massachusetts and the Infrared Processing and Analysis 
Center/California Institute of Technology, funded by the National 
Aeronautics and Space Administration and the National Science Foundation.}
\end{center}
\end{figure}

\section{$FUSE$ Observations and Analysis}

The diffuse region of the Orion Nebula located at RA~=~05$^{h}$35$^{m}$12.43$^{s}$, 
$\delta$~=~-05$^{\circ}$11\arcmin35.3\arcsec, J2000, was observed 
by $FUSE$ for 16.7 ks in time-tagged (TTAG) mode on 2001 November 26 as part of the S405 channel realignment
program described by~\citet{murthy04}~and~\citet{murthy05}.  
Nebular spectra were acquired across the $FUSE$ bandpass (905~--~1187~\AA) 
in the (30\arcsec~$\times$~30\arcsec) LWRS aperture (Figure 1). 
Descriptions of the $FUSE$ instrument and on-orbit performance 
characteristics can be found in Moos et al. (2000) and Sahnow et al. (2000).~\nocite{moos00,sahnow00} 
The spectra were obtained from the Multi-Mission Archive at STScI (MAST),  
and were analyzed with the original 
CalFUSE calibration (version 1.8.7). The calibrated data files for each orbit
were then coadded using IDL software and the individual channels
were combined using a cross-correlation algorithm.
The exact pointing was determined by comparing Fine Error Sensor (FES) 
images with images generated by sky simulator software.  

We present the S4054601 spectra at
wavelengths longer than Ly-$\beta$ (Figure 2), where the bulk of the far-UV emission
from H$_{2}$ resides.  In addition to the numerous absorption lines coming from 
molecular hydrogen discussed below, we see the scattered photospheric lines
from \ion{C}{3}~$\lambda$977/1176, \ion{P}{5}~$\lambda$1118/1128, and \ion{Si}{4}~$\lambda$1122/1128, 
as well as interstellar absorption from \ion{C}{2}~$\lambda$1036 and
\ion{N}{1}~$\lambda$1135.  These features are identified in Figure 2.
The fact that the \ion{P}{5}~$\lambda$1118
line was seen in the spectrum is probably the most conclusive argument
for $\theta^{1}$ Ori C (as opposed to HD 36981) illuminating this region of Orion.  
Atomic hydrogen lines are highly contaminated by the interstellar medium and~\citet{pellerin02} 
find that \ion{P}{5} is not present in stars
of spectral type later than B2V.  
Finally, we find an unidentified absorption feature
at 1031.4~\AA\ that we tentatively assign to blue-shifted \ion{O}{6}.
Blue-shifted \ion{O}{6} absorption is seen in many stars of of this spectral
type~\citep{pellerin02}, 
however we cannot conclusively identify it as the $\lambda$1038 component
of the doublet falls in a region dominated by H$_{2}$ and 
\ion{C}{2} absorption.

\section{Modeling Molecular Hydrogen in Orion} \label{model}

The ultraviolet absorption and emission features of H$_{2}$ observed in 
Orion are seen imposed on the dust scattered continuum of 
$\theta^{1}$ Ori C~\citep{murthy05}.  A model of the nebular dust-scattering continuum is beyond
the scope of this work, but we set the baseline for the molecular 
hydrogen by fitting the continuum with a double power-law of the form
\begin{equation}
S(\lambda)~=~(\frac{\lambda}{\lambda_{0}})^{\alpha}~-~C(\frac{\lambda-\lambda_{0}}{\lambda_{0}})^{\beta}
\end{equation}
scaled to the data.  This simple emperical expression with $\lambda_{0}$ of 
1187~\AA, yields $\alpha$~=~0.1, $\beta$~=~2.0, and $C$~=~5.0.  An H$_{2}$
absorption spectrum was calculated using  the $H_{2}ools$ optical depth 
templates~\citep{h2ools}.  As the Doppler $b$-value  
is uncertain, we have adopted 2 km s$^{-1}$ for this diffuse region, 
consistent with $b$-values observed along the line of sight to other
stars in this region of Orion~\citep{spitzer74,jenkins97}. 
With that assumption, we 
used the templates to fit the (4~-~0) band ($\lambda~\approx$~1050~\AA)
and find a  total  $N(H_{2})$~=~4.7~$\times$~10$^{18}$~cm$^{-2}$.
Absorption out of the 
ground electro-vibrational state for the first six rotational states
was considered, but most of the absorption comes from 
$j$~=~0 and 1.  
Because the absoption lines are partially filled in by the 
fluorescent emission (see below), the true $N(H_{2})$ for this 
line of sight is most likely greater than 4.7~$\times$~10$^{18}$~cm$^{-2}$.
This level of H$_{2}$ absorption is surprising, as studies of stars in this 
region find a very low molecular fraction
in the intervening gas (Abel et al. 2004 and references therein).~\nocite{abel04} 

We fit the emission spectrum of H$_{2}$ seen in these $FUSE$ data using the 
hydrogen fluorescence model described in~\citet{france05}.
Using parameters for the Orion region found in the literature 
for $N(H_{2})$ and temperature, 
this fluorescent emission model assumes a ground electronic
state population, then uses photoexcitation cross sections and 
an incident radiation field to calculate the rovibrational levels of 
the upper electronic states (predominantly $B$$^{1}\Sigma^{+}_{u}$~and~$C$$^{1}\Pi_{u}$).
The molecules then return to the ground electronic state following
the appropriate selection rules and branching ratios, producing
the observed ultraviolet emission lines and leaving the molecules
in excited rovibrational levels.  \citet{kristensen03} imaged
regions of the Orion Nebula in the near-IR rovibrational lines 
of H$_{2}$ to determine the excitation temperature.  
They find a clear 
delineation between hot and cold zones, requiring more than one temperature
component to produce a reasonable fit to the observations.  We therefore adopted 
the rotational temperature of 390 K from~\citet{habart04} as 
the cool (thermal) component and a vibrational temperature of 2500 K as representative 
of the hot (shocked or non-thermal) component~\citep{france05}.
An $N(H_{2})$~=~1~$\times$~10$^{21}$~cm$^{-2}$~\citep{habart04} was used with a 
$b$-value of 2 km s$^{-1}$.  Each of these parameters
affects the resultant model spectrum differently. 
The model spectra change very little with $b$ for values of a few km s$^{-1}$. 
Varying the column density changes the total emitted power, higher
column density giving more output photons, but self-absorption begins to 
surpress discrete lines at columns greater than~$\sim$~5~$\times$~10$^{21}$
cm$^{-2}$.  Temperature controls both the shape and scale of the
model spectra, with more levels of the ground electronic state are populated
at higher temperatures.  

\begin{figure*}
\begin{center}
\epsscale{0.75}
\rotatebox{90}{
\plotone{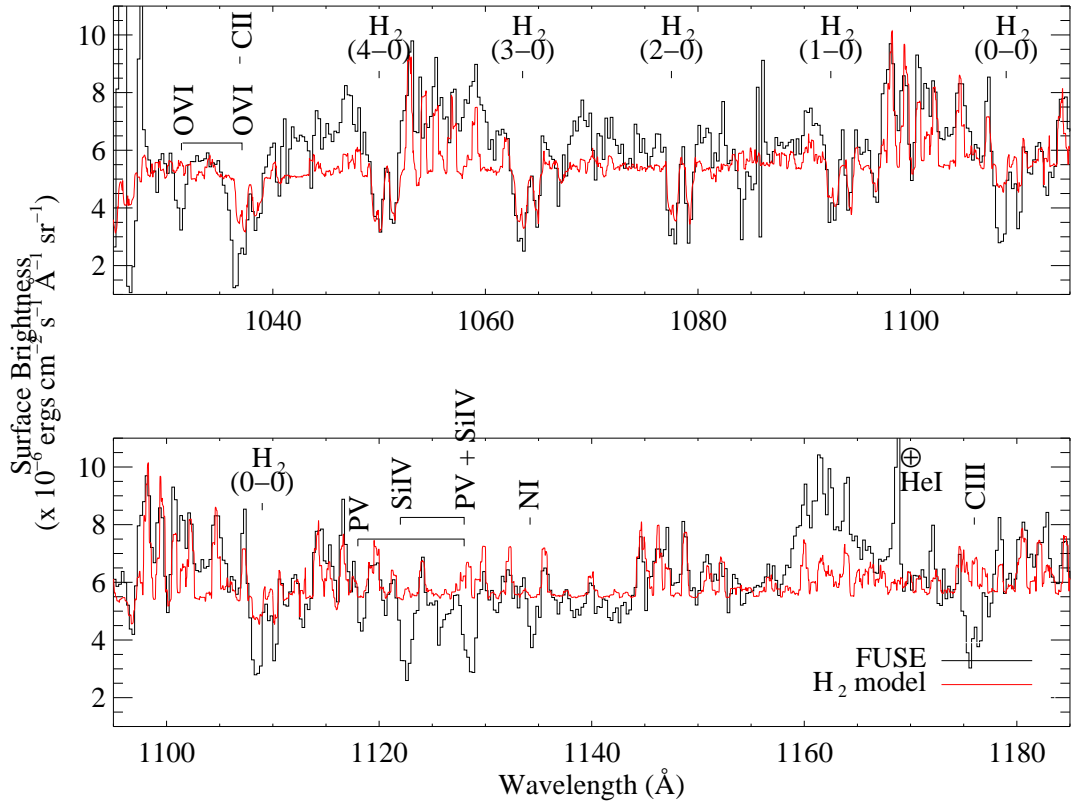} }
\caption{\label{fuseneb}  The $FUSE$ spectrum described in the text overplotted 
with the H$_{2}$ emission model described in \S 3.  The model spectrum
has been convolved to a resolution of 0.5~\AA.  H$_{2}$ lines are the dominant features
seen in both absorption and emission between Ly-$\beta$ and the end of the 
$FUSE$ bandpass.  \ion{C}{2} and \ion{N}{1} are of interstellar origin.
The longest wavelength H$_{2}$ absorption strengths
are underpredicted due to an offsetting contribution from the emission lines, 
as described in \S~4.}
\end{center}
\end{figure*}

Radiation from $\theta^{1}$ Ori C dominates the ambient field in this part of
Orion~\citep{murthy05}.  It is most likely responsible for pumping the
observed fluorescence, and should thus be used as the input field for modeling
the H$_{2}$ emission.
This is complicated by two factors: 1) a well flux-calibrated far-UV
spectrum of $\theta^{1}$ Ori C does not exist and 2)
the majority of the absorption seen on the HD 37022 line of sight is thought
to be located in intervening clouds~\citep{bautista95,murthy05}.
A spectrum of $\theta^{1}$ Ori C as observed from Earth will not be representative
of that seen by the molecules in Orion. 
In an attempt to overcome these concerns, 
we chose a lightly reddened SMC star of similar spectral type
(AV 243, O6V, $E_{B~-~V}$~=~0.09) from the 
$FUSE$ Magellenic Cloud Atlas of Danforth et al (2002; see also
Walborn et al. 2002).~\nocite{danforth02,walborn02} 
A simple correction for the differences between Milky Way and SMC
dust was made using the computed extinction curves of~\citet{wd01}, 
assuming the reddening of $\theta^{1}$ Ori C at the $FUSE$ pointing
is similar to what is observed along the HD~36981 sightline
($E_{B~-~V}$~=~0.04).  We assumed that the reddening
on the AV 243 line of sight was due to equal contributions from 
Milky Way ($R_{V}$~=~5.5) and SMC dust.  An $IUE$ spectrum of
$\theta^{1}$ Ori C (swp01394, also obtained from MAST; Bohlin \& Savage 1981)
was used to cross-calibrate the surrogate $FUSE$ spectrum in the 
overlap region between 1150 and 1187~\AA.~\nocite{bohlin81}
Primary uncertainties in this approach are 
a lack of knowledge about the attenuation between the star and the 
nebula, the actual contribution from SMC and Milky Way dust, 
and possible calibration problems with the $IUE$ spectrum~\citep{fm00}.   
Using this incident radiation field and the parameters 
described above, we constructed the fluorescence 
spectrum (convolved with a 0.5~\AA\ box kernel) for
comparison with the data.

\section{Discussion}

The nebular spectra and the model described above are shown in Figure 2.
The model finds satisfactory qualitative agreement with the relative line strengths seen 
in the $FUSE$ spectra between Ly-$\beta$ and the end of the bandpass.
We find effects similar to those seen in far-UV spectra of the 
emission/reflection nebula IC 63, namely
that the observed H$_{2}$ emission lines are seen to be broader than 
what is expected from the instrumental profile for these filled aperture 
observations~\citep{france05}.  A study of the possible mechanisms that would produce this
broadening is underway.  The two major discrepancies that we
find between the data and the relative strengths  predicted by the model
are in the region between 1040~--~1050~\AA, and the band of lines centered on 1161~\AA.  
The 1040~--1050 region may be contaminated by broad wind emission from 
\ion{O}{6}, as $\theta^{1}$ Ori C also shows an ``excess'' in this range~\citep{murthy05}, 
but the exact cause is uncertain.  The 1161 band seems to be
sitting on a ``shelf'' that is not predicted by the model.  One hypothesis that
we explored was the possibility of hot H$_{2}$ (T~$\geq$~2500 K) being
fluoresced by nebular Ly-$\alpha$, as is seen in the far-UV spectra
of pre-main-sequence stars~\citep{wilkinson02} and accreting binary systems~\citep{wood04}.
In order to test this theory, we added a linear continuum that extended the
excitation spectrum to 1300~\AA\ and a Gaussian emission line
at Ly-$\alpha$.  This additional excitation out of excited states of H$_{2}$
does not resolve our problem with the 1161 band.  We find that as the 
1161 band begins to fill in that numerous other lines are predicted that
we do not observe in the $FUSE$ spectra.  The most likely
conclusion is that our model is lacking some physical mechanism that
gives rise to these features.

The model does not agree with the observed absolute flux.  
We find that in order to correctly predict the observed absolute flux, 
our model output needs to be scaled up by a factor of two.  We attribute this
to three possible causes: 1) an uncertainty in the degree of H$_{2}$ clumping
in this region, 2) an uncertainty of the true strength of the exciting
radiation field, and 3) the (unknown) mechanism that
causes discrepancies at 1045 and 1161~\AA.  
Dust and molecular hydrogen are known to be found
in clumpy and/or filamentary structures in regions where hot stars interact
with their surroundings~\citep{odell00,kristensen03,france04}, however we
find this explanation unlikely. Creating similar models at higher 
column densities shows that while the overall emitted power increases
with column, the majority of this extra emission 
is located outside of the $FUSE$ bandpass ($\lambda~\geq$ 1200~\AA). 
The second possibility is that our input excitation
field does not accurately represent the local absolute 
radiation field.  We feel confident that the spectral characteristics
of the input spectrum are good based on the relative agreement with the H$_{2}$ emission 
spectrum, but are less confident in the absolute scale.  
Changes in our reddening assumptions, absolute calibration of the $IUE$ spectrum
we used~\citep{fm00}, and uncertainties in the 
nebular distance/geometry can easily explain the scaling required. 
Finally, further modeling efforts are underway to resolve the problems described above.

The molecular hydrogen absorption lines present in the spectra also show
an interesting behavior.  One would expect, based on the decreasing
oscillator strengths~\citep{h2ools}, that the lines would be shallower during the
progression to smaller vibrational transitions (4~--~0 to 0~--~0, 
shown in Figure 2).   We find that the absorption profiles
are of almost identical depth in each of the bands.  We attribute this
to a filling effect by the emission from H$_{2}$ where the emitting and 
absorbing transitions overlap.  There are a greater number of 
emitting transitions overlapping with the absorption lines at shorter wavelengths, 
so fitting the absorption lines at the (4~--~0) band
leads to an under prediction of the strength of the lines
in the longer wavelength bands of molecular hydrogen.  
This finding leads us to stress that meaningful column densities cannot
be determined without correcting for the emission lines and 
properly addressing the dust scattering to set the continuum level.

There has been uncertainty in the degree to which shocks or ultraviolet
photo-excitation leads to the observed near-IR lines of molecular
hydrogen seen in Orion~\citep{burton89,luhman94,kristensen03}.  
These $FUSE$ data rule out models that consider only shocks as the mechanism 
producing the near-IR emission.  Shocks cannot populate the excited
electronic states of molecular hydrogen that give rise to the 
emission lines seen~\citep{shull82}.  Hydrogen molecules will be
dissociated by collisions at temperatures greater than roughly
4000~K, thus they would be destroyed before reaching temperatures
sufficient to begin populating the first excited electronic state, 
$B$$^{1}\Sigma^{+}_{u}$ (E~$\approx$~11 eV).  These data cannot
exclude shocks from making a contribution to the near-IR H$_{2}$
emission~\citep{kristensen03}, but ultraviolet excitation
clearly contributes to the rovibrationally excited population 
of hydrogen molecules in the Orion Nebula.

\acknowledgments
We would like to acknowledge Jayant Murthy and Dave Sahnow for alerting 
us to the existence of this $FUSE$ data and helpful discussion.
We thank Paul Feldman for his insights on molecular hydrogen and
Alex Fullerton for enjoyable discussion about the ultraviolet spectra of hot stars.
This paper has been greatly improved by the comments of an anonymous referee.
Observations were obtained by the 
NASA-CNES-CSA $FUSE$ mission, operated by the Johns Hopkins University.

%%%%%%%%%%%%%%%%%%%%%%%%%%%BIBLIOGRAPHY%%%%%%%%%%%%%%%%%%
\bibliography{ms}

\begin{thebibliography}{27}
\expandafter\ifx\csname natexlab\endcsname\relax\def\natexlab#1{#1}\fi

\bibitem[{{Abel} {et~al.}(2004){Abel}, {Brogan}, {Ferland}, {O'Dell}, {Shaw},
  \& {Troland}}]{abel04}
{Abel}, N.~P., {Brogan}, C.~L., {Ferland}, G.~J., {O'Dell}, C.~R., {Shaw}, G.,
  \& {Troland}, T.~H. 2004, \apj, 609, 247

\bibitem[{{Bautista} {et~al.}(1995){Bautista}, {Pogge}, \&
  {Depoy}}]{bautista95}
{Bautista}, M.~A., {Pogge}, R.~W., \& {Depoy}, D.~L. 1995, \apj, 452, 685

\bibitem[{{Bohlin} \& {Savage}(1981)}]{bohlin81}
{Bohlin}, R.~C. \& {Savage}, B.~D. 1981, \apj, 249, 109

\bibitem[{{Burton} {et~al.}(1989){Burton}, {Brand}, {Geballe}, \&
  {Webster}}]{burton89}
{Burton}, M.~G., {Brand}, P.~W.~J.~L., {Geballe}, T.~R., \& {Webster}, A.~S.
  1989, \mnras, 236, 409

\bibitem[{{Danforth} {et~al.}(2002){Danforth}, {Howk}, {Fullerton}, {Blair}, \&
  {Sembach}}]{danforth02}
{Danforth}, C.~W., {Howk}, J.~C., {Fullerton}, A.~W., {Blair}, W.~P., \&
  {Sembach}, K.~R. 2002, \apjs, 139, 81

\bibitem[{{France} {et~al.}(2005){France}, {Andersson}, {McCandliss}, \&
  {Feldman}}]{france05}
{France}, K., {Andersson}, B.-G., {McCandliss}, S.~R., \& {Feldman}, P.~D.
  2005, \apj

\bibitem[{{France} {et~al.}(2004){France}, {McCandliss}, {Burgh}, \&
  {Feldman}}]{france04}
{France}, K., {McCandliss}, S.~R., {Burgh}, E.~B., \& {Feldman}, P.~D. 2004,
  \apj, 616, 257

\bibitem[{{Habart} {et~al.}(2004){Habart}, {Boulanger}, {Verstraete},
  {Walmsley}, \& {Pineau des For{\^ e}ts}}]{habart04}
{Habart}, E., {Boulanger}, F., {Verstraete}, L., {Walmsley}, C.~M., \& {Pineau
  des For{\^ e}ts}, G. 2004, \aap, 414, 531

\bibitem[{{Jenkins} \& {Peimbert}(1997)}]{jenkins97}
{Jenkins}, E.~B. \& {Peimbert}, A. 1997, \apj, 477, 265

\bibitem[{{Kristensen} {et~al.}(2003){Kristensen}, {Gustafsson}, {Field},
  {Callejo}, {Lemaire}, {Vannier}, \& {Pineau des For{\^ e}ts}}]{kristensen03}
{Kristensen}, L.~E., {Gustafsson}, M., {Field}, D., {Callejo}, G., {Lemaire},
  J.~L., {Vannier}, L., \& {Pineau des For{\^ e}ts}, G. 2003, \aap, 412, 727

\bibitem[{{Luhman} {et~al.}(1994){Luhman}, {Jaffe}, {Keller}, \&
  {Pak}}]{luhman94}
{Luhman}, M.~L., {Jaffe}, D.~T., {Keller}, L.~D., \& {Pak}, S. 1994, \apjl,
  436, L185

\bibitem[{{Ma{\'{\i}}z-Apell{\' a}niz} {et~al.}(2004){Ma{\'{\i}}z-Apell{\'
  a}niz}, {Walborn}, {Galu{\' e}}, \& {Wei}}]{maiz04}
{Ma{\'{\i}}z-Apell{\' a}niz}, J., {Walborn}, N.~R., {Galu{\' e}}, H.~{\' A}.,
  \& {Wei}, L.~H. 2004, \apjs, 151, 103

\bibitem[{{Massa} \& {Fitzpatrick}(2000)}]{fm00}
{Massa}, D. \& {Fitzpatrick}, E.~L. 2000, \apjs, 126, 517

\bibitem[{{McCandliss}(2003)}]{h2ools}
{McCandliss}, S.~R. 2003, \pasp, 115, 651

\bibitem[{{Moos}(2000)}]{moos00}
{Moos}, H.~W. et~al. 2000, \apjl, 538, L1

\bibitem[{{Murthy} \& {Sahnow}(2004)}]{murthy04}
{Murthy}, J. \& {Sahnow}, D.~J. 2004, \apj, 615, 315

\bibitem[{{Murthy} {et~al.}(2005){Murthy}, {Sahnow}, \& {Henry}}]{murthy05}
{Murthy}, J., {Sahnow}, D.~J., \& {Henry}, R.~C. 2005, \apjl, 618, 99

\bibitem[{{O'Dell}(2000)}]{odell00}
{O'Dell}, C.~R. 2000, \aj, 119, 2311

\bibitem[{{Pellerin} {et~al.}(2002){Pellerin}, {Fullerton}, {Robert}, {Howk},
  {Hutchings}, {Walborn}, {Bianchi}, {Crowther}, \& {Sonneborn}}]{pellerin02}
{Pellerin}, A., {Fullerton}, A.~W., {Robert}, C., {Howk}, J.~C., {Hutchings},
  J.~B., {Walborn}, N.~R., {Bianchi}, L., {Crowther}, P.~A., \& {Sonneborn}, G.
  2002, \apjs, 143, 159

\bibitem[{{Rachford} {et~al.}(2002){Rachford}, {Snow}, {Tumlinson}, {Shull},
  {Blair}, {Ferlet}, {Friedman}, {Gry}, {Jenkins}, {Morton}, {Savage},
  {Sonnentrucker}, {Vidal-Madjar}, {Welty}, \& {York}}]{rachford02}
{Rachford}, B.~L., {Snow}, T.~P., {Tumlinson}, J., {Shull}, J.~M., {Blair},
  W.~P., {Ferlet}, R., {Friedman}, S.~D., {Gry}, C., {Jenkins}, E.~B.,
  {Morton}, D.~C., {Savage}, B.~D., {Sonnentrucker}, P., {Vidal-Madjar}, A.,
  {Welty}, D.~E., \& {York}, D.~G. 2002, \apj, 577, 221

\bibitem[{{Sahnow}(2000)}]{sahnow00}
{Sahnow}, D.~J. et~al. 2000, \apjl, 538, L7

\bibitem[{{Shull} \& {Beckwith}(1982)}]{shull82}
{Shull}, M. \& {Beckwith}, S. 1982, Annual Reviews of Astronomy and
  Astrophysics, 30, 163

\bibitem[{{Spitzer} {et~al.}(1974){Spitzer}, {Cochran}, \&
  {Hirshfeld}}]{spitzer74}
{Spitzer}, L., {Cochran}, W.~D., \& {Hirshfeld}, A. 1974, \apjs, 28, 373

\bibitem[{{Walborn} {et~al.}(2002){Walborn}, {Fullerton}, {Crowther},
  {Bianchi}, {Hutchings}, {Pellerin}, {Sonneborn}, \& {Willis}}]{walborn02}
{Walborn}, N.~R., {Fullerton}, A.~W., {Crowther}, P.~A., {Bianchi}, L.,
  {Hutchings}, J.~B., {Pellerin}, A., {Sonneborn}, G., \& {Willis}, A.~J. 2002,
  \apjs, 141, 443

\bibitem[{{Weingartner} \& {Draine}(2001)}]{wd01}
{Weingartner}, J.~C. \& {Draine}, B.~T. 2001, ApJ, 548, 296

\bibitem[{{Wilkinson} {et~al.}(2002){Wilkinson}, {Harper}, {Brown}, \&
  {Herczeg}}]{wilkinson02}
{Wilkinson}, E., {Harper}, G.~M., {Brown}, A., \& {Herczeg}, G.~J. 2002, \aj,
  124, 1077

\bibitem[{{Wood} \& {Karovska}(2004)}]{wood04}
{Wood}, B.~E. \& {Karovska}, M. 2004, \apj, 601, 502

\end{thebibliography}


%%%%%COPYING THE FIGURE INPUT FROM PAUL's IO TORUS PAPER, 7/16/03%%%%%

%% If you are not including electronic art with your submission, you may
%% mark up your captions using the \figcaption command. See the 
%% No more than seven \figcaption commands are allowed per page, 
%% so if you have more than seven captions, insert a \clearpage 

%% The following command ends your manuscript. LaTeX will ignore any text

\end{document}